\documentstyle[12pt]{article}
\begin{document}
\title{Guage conditions for an Abelian Chern-Simons
system consistent with equations of motion}
\author{Pankaj Sharan\thanks{email : pankj.ph@jmi.ernet.in},
Anupama Mehra,
 Krishnendu Dasgupta\thanks{email: kd@arbornet.org} \\ \and M. Sami
\\ Department of Physics, Jamia Millia Islamia\\New Delhi 110025}
\maketitle
\abstract{Complete constraint analysis and choice of gauge
 conditions consistent with equations of motion is done for
  Abelian Chern Simons field interacting minimally with a
   complex scalar field.The Dirac-Schwinger consistency condition is
   satisfied by the reduced phase space Hamiltonian density
   with respect to the the Dirac bracket.It is shown
   that relativistic invariance
   under boosts can be obtained only if gauge conditions were chosen
   consistent with the equations of motion.Moreover all gauge invariant
   quantities are shown to be free of transformation anomaly. }

\section{Introduction}
Canonical formalism for Abelian Chern-Simons field system
interacting a  scalar field in $2+1$
dimensions has been discussed by many authors$^{1-5}$in connection
with particles obeying fractional statistics or
 ``anyons".

This system is the fundamental model for relativistic anyons,and its
quantization reveals many intricate points concerning constrained
field systems.Anyons are of physical relevance to phenomena like
fractional quantum Hall effect and possibly to high $T_C$ superconductivity
$^5$.

Despite its importance, the field theory of this system remains incomplete
. One reason is the lack of explicit locality in the reduced
phase space. Another obstacle has been the lack
 of a set of gauge conditions consistent with equations of motion.
Although different gauge conditions have been discussed,
the authors have not considered consistency with equations of motion. As we
see in section $4$, the requirement is essential for
the system to be relativistically
invariant. In particular, transformations under boosts ( which implicitly
involve equations of motion) must lead to proper transformations of
the gauge field $A^\mu$ to make the system Lorentz frame independent.

 In this paper we do the complete constraint analysis of the system.
 There are two second class and two first class constraints. We write down
equations of motion,and then use the freedom of gauge transformations to
guess the gauge conditions consistent with the equations of motion.
We then proceed to remove extra variables by constructing Dirac brackets with
the two
second class constraints and the two first class ones and their corresponding
gauge conditions treated as two pairs of second class constraints.
We do the construction of Dirac brackets in steps,using a pair at each step.
The reduced phase space Hamiltonian surprisingly has the same form as one
obtained earlier by using wrong gauge conditions.$ ^2$

We then discuss the Dirac-Schwinger consistency
conditions to obtain stess energy tensor density $\Theta^{0i}$.The Poincare
generators
are constructed , and\\ `` anomalous" transformation properties of the basic
fields are discussed.It is shown that the conserved current obeys normal
transformation laws as a 2+1 vector.Importance of boost transformations
to gaurantee relativistic invariance is emphasized.

The present work completes the first stage of writing a canonical formalism
for the Abelian Chern-Simons system.The quantization and attending
operator ordering and locality properties will be discussed
in a separate communication.
\newcommand{\k}{\kappa}
\newcommand{\dd}{\delta}
\newcommand{\oo}{\over}
\newcommand{\x}{\vec x}
\newcommand{\y}{\vec y}
\newcommand{\lr}[1]{\stackrel{\leftrightarrow}{#1}}
\newcommand{\p}{\partial}
\newcommand{\e}{\epsilon}
\newcommand{\n}{\nabla}
\newcommand{\jj}{\jmath}
\newcommand{\pp}{\phi}
\newcommand{\lrr}{\Longrightarrow} 
\section{Constraint analysis and equations of motion}
Our $2+1$ dimensional system is described by the Lagrangian density

$${\cal L}=(D_\mu \phi)^*(D^\mu \phi)-m^2\pp^*\pp+{\k\oo 2}
\e_{\mu \nu \rho}A^\mu\p^\nu A^\rho \eqno{(2.1)}$$
where
$$D_\mu=\p_\mu-ieA_\mu \eqno{(2.2)}$$
The canonical momenta are defined by
 $$\pi_\mu={1\oo 2}\k\e_{0 \mu \sigma}A^\sigma \eqno{(2.3)}$$
$$ p=(D_0 \pp)^*=\dot{\pp^*}+ieA_0 \pp^* \eqno{(2.4)}$$
$$ p^*=(D_0 \pp)=\dot\pp-ieA_0 \pp \eqno{(2.5)}$$
All the three momenta canonical to $ A^\mu$ give rise to three primary
constraints.
$$\Phi_0 \equiv \pi_0\approx 0\eqno{(2.6)}$$
$$\Phi_1 \equiv \pi_1-{1\oo 2}\k A^2 \approx 0\eqno{(2.7)}$$
$$\Phi_2 \equiv \pi_2+{1 \oo 2} \k A^1 \approx 0\eqno{(2.8)}$$
The Hamiltonian can be written as
\begin{eqnarray*}
H&=&\int d^2 x [p^*p+A^0(\k \vec \n \times \vec A-\jj^0)
+\vec \n \pp^*\cdot\vec \n \pp \\
& &\qquad -ie \vec A \cdot (\pp^* \lr{ \n} \pp)
+e^2 |\vec A |^2 |\pp|^2 +m^2 |\pp|^2] \\
&+& \int d^2 x (v^0 \Phi_0+v^1 \Phi_1+v^2 \Phi_2)
\end{eqnarray*}                                $$ \eqno{(2.9)}$$
    where $$ \vec A=(A^1,A^2)\qquad,\qquad\vec \n=(\p_1,\p_2)
    \qquad,\qquad\vec \n \times \vec A=\p_1 A^2- \p_2 A^1$$ and
$$ \jj^0= ie (p^*\pp^*- p\pp)=ie(\pp^* D_0 \pp -(D_0 \pp)^* \pp)\eqno{(2.10)}$$
and  $v^0,v^1,v^2$ are ``\,undetermined" velocities or Lagrangian multipliers.
Using the canonical Poisson brackets
$$\{\pp(\x,t),p(\y,t)\}=\dd(\x-\y)\eqno{(2.11)} $$
$$\{\pp^*(\x,t),p^*(\y,t)\}=\dd(\x-\y)\eqno{(2.12)}$$
$$\{A^\mu(\x,t),\pi_\nu(\y,t)\}=\dd^\mu_\nu \dd(\x-\y)\eqno{(2.13)}$$
we can enquire into the permanence of the primary constraints

$$\{\Phi_1,H\}=0\eqno{(2.14)}$$
$$\{\Phi_2,H\}=0\eqno{(2.15)}$$
which determine the velocities
$$v^1=-\p_1 A^0 -{1 \oo \k} j_2\eqno{(2.16)}$$
$$v^2=-\p_2 A^0 +{1 \oo \k} j_1\eqno{(2.17)}$$
where
$$\jj_i  =  ie(\pp^* \lr \p_i \pp)-2e^2 A^i |\pp|^2
      =  ie(\pp^* D_i \pp-(D_i \pp)^*\pp)\eqno{(2.18)}$$
are the space components of the current
$\jj_\mu=ie \pp^*\lr{D_\mu} \pp\,$where we use the notation $\pp^*\stackrel
\leftarrow D_\mu=(D_\mu \pp)^*$ .
The pair $\Phi_1,\Phi_2$ is obviously
second class.

The condition for permanence of the constraint$\,\Phi_0\approx 0$
 gives a secondary
constraint$$\{\Phi_0,H\}=0 \lrr
\Phi_3\equiv \k \vec \n \times \vec A -\jj^0\approx 0\eqno{(2.19)}$$
which can be made a first class constraint by taking the linear combination
$$\Psi\equiv\Phi_3+\p_1 \Phi_1 +\p_2 \Phi_2 =
\vec \n\cdot\vec \Pi+{\k\oo 2}\vec \n \times \vec A-\jj^0\approx 0
\eqno{(2.20)}$$
where $\vec \Pi=(\pi_1,\pi_2)$.  \\   \\
As$\, \{\Phi_3,H\} $ vanishes due to earlier constraints there are no further
constraints in the system.
After substituting $v^1$ and$\,v^2$ the Hamiltonian can be written as\\
\begin{eqnarray*}
H & = & \int d^2 x [p^* p +|\vec \n \pp |^2 + m^2 |\pp |^2]   \\
  &+ & \int d^2 x [{ ie\oo \k }(\pp^* \lr{ \p_1 } \pp)(\pi_2-{1\oo 2}\k A^1)
-{ ie\oo \k }(\pp^* \lr{ \p_2 } \pp)(\pi_1+{1\oo 2}\k A^2 )\\
&  & \qquad \qquad   
-{2e^2\oo \k} |\pp|^2(A^1 \pi_2-A^2 \pi_1)]\\
  & + & \int d^2 x [v^0 \Phi_0 + A^0 \Psi ]
\end{eqnarray*} $$\eqno{(2.21)} $$
Equations of motion can now be written 
$$\dot\pp=\{\pp,H\}= p^* +ieA_0 \pp\eqno{(2.22)}$$
$$\dot\pp^*=\{\pp^*,H\}=p -ieA_0 \pp^*\eqno{(2.23)}$$
$$\dot p =\{ p,H \}=(\vec D\cdot \vec D \pp)^*-m^2 \pp^*
-ieA_0 p\eqno{(2.24)}$$
$$\dot p^* =\{ p^*,H \}=\vec D\cdot \vec D \pp-m^2 \pp
+ieA_0 p^*\eqno{(2.25)}$$
$$\dot A_0=v^0\eqno{(2.26)}$$
$$ \dot A^1=-\p_1 A^0-{1\oo \k} j_2= v^1\eqno{(2.27)}$$
$$ \dot A^2=-\p_2 A^0+{1\oo \k} j_1= v^2 \eqno{(2.28)}$$
As expected, the ``undetermined "  velocity $v^0$ is still undetermined
because of the first class constraint $\Phi_0$.This makes the evolution of
$ A_0 \quad$completely arbitrary The ``\,determined" velocities $v^1,v^2$ are
only partially determined as there is freedom in $ A_0 $.

Before turning to gauge conditions and reduced phase space
via Dirac brackets, we note that, as a consequence of equations of
motion the current $\jj_\mu=ie \pp^*\lr{D_\mu} \pp\quad$ is conserved.
$$\p^\mu \jj_\mu=0\eqno{(2.29)}$$
\section{Gauge conditions and Reduced Phase space}
We have to choose gauge conditions which have non-zero Poisson brackets
with the two first class constraints $\Phi_0 $ and $\Psi $,and which also
remain satisfied in time. the obvious choice,based on the familiar case
of electrodynamics is to use freedom in evolution of $ A_0 $ such that
we can fix $\vec \n\cdot \vec A=0$ as a gauge condition for
constraint $\Psi $ which involves $\vec \n\cdot \vec \Pi $.

We see from $(2.27,2.28)$ that
$${d\oo dt}\vec \n\cdot \vec A=-\n^2 A^0-{1\oo \k}(\p_1 \jj_2-\p_2\jj_1)
\eqno{(3.1)}$$
Therfore$\,\vec \n\cdot \vec A$ will remain zero at all times if $A^0$
is chosen to be such that
$$\chi_0\equiv A^0(x) +{1\oo \k }\int d^2 y G(\x-\y)(\p_1\jj_2-\p_2\jj_1)
=0\eqno{(3.2)}$$
where $ G(\x-\y)$ is the solution of
$$\n^2 G(\x-\y)=\dd(\x-\y)\eqno{(3.3)}$$
equal to $$G(\x-\y)={1\oo2 \pi}\ln(|\x-\y|)\eqno{(3.4)}$$

We therefore choose $ \chi_0$ as our first gauge condition corresponding
to the first class constraint $\Phi_0\approx 0$,
and
$$\chi\equiv \vec \n\cdot \vec A=0\eqno{(3.5)}$$
as the second gauge condition corresponding to the first
 class constraint $\Psi\approx 0$.

Thus our system has the following pairs of conditions
$$\Phi_1,\Phi_2\quad;\quad \Phi_0,\chi_0\quad;\quad\Psi,\chi \eqno{(3.6)}$$
which can be treated like a set of three pairs of second class constraints.

The Dirac brackets can be constructed most conveniently
 by eliminating extra deegrees of freedom step by step forming
new brackets from old ones by incorporating one
 pair at a time.After forming the  Dirac bracket, the conditions can be
 strongly put equal to zero,eliminating extra degrees of freedom.

 The Poisson bracket matrix $C_{1ij}=\{\Phi_i,\Phi_j\}$ where$ (i,j=1,2)$ is

 $$C_{1ij}(\x,\y)=\left(
 \begin{array}{cc}
 0          & -\k\dd(\x-\y)      \\
\k\dd(\x-\y) &   0
\end{array}
\right)
$$
$$\eqno(3.7)$$
therefore the Dirac bracket
$$\{ A, B\}_1 = \{A,B\} - \int d^2 x \int d^2 y
\{ A,\Phi_i(\x,t)\}\,C^{-1}_{1ij}(\x,\y)\,\{\Phi_j(\y,t), B\}$$
$$\eqno(3.8)$$
gives the basic brackets $\{\,,\,\}_1$  which are the same as
canonical Poisson brackets except that $\pi_1,\pi_2$ are eliminated
as they can be put equal to ${1 \oo 2}\k A^2 $ and $-{1 \oo 2}\k A^1$
respectively
now,with the consequence that
$$\{ A^1, A^2\}_1 ={1\oo \k}\dd(\x-\y)\eqno{(3.9)}$$
The remaining constraint or gauge conditions are at this stage 
$$\Phi_0=0\eqno{(3.10)}$$
$$\chi_0=A_0+{ 1\oo \k}\int d^2 y G(\x-\y)(\p_1 \jj_2-\p_2 \jj_1)(y)=0
\eqno{(3.11)}$$
$$\Psi=\k (\vec \n \times \vec A)-\jj^0=0\eqno{(3.12)}$$
$$\chi=\vec \n \cdot \vec A=0\eqno{(3.13)}$$
and the reduced Hamiltonian
\begin{eqnarray*}
H & = & \int d^2 x [ p^* p +|\vec \n \pp|^2+m^2|\pp|^2]   \\
 & + & \int d^2 x [-ie \vec A \cdot(\pp^*\lr{\n}\pp)+e^2|\pp|^2|\vec A |^2]\\
 & + & \int d^2 x [v^0 \Phi_0+A^0\Psi]
\end{eqnarray*}                        $$\eqno{(3.14)} $$

Now we form the matrix $C_{2ij} $ from $\Phi_0,\chi_0 $
 which again has trivial $\dd $ functions as the off-diagnol elements,
  and the
 Dirac bracket$\{\, ,\,\}_2$ constructed from $\{\,,\,\}_1$
using $ C^{-1}_2 $ gives no changes in the brackets except that $\Phi_0 $
and $\chi_0$ can be put strongly equal to zero eliminating $\pi_0$ and $ A_0$
from the dynamics with $ A_0 $ given by $ (3.11) $.

As the final step we form Dirac bracket with $\Psi$ and $\chi$ .
The bracket matrix is
$$C_3=\left(
\begin{array}{cc}
0               & -\n^2 \dd(\x-\y) \\
\n^2 \dd(\x-\y) &   0
\end{array}
\right)$$
$$ \eqno{(3.15)}$$
whose inverse
$$C_3^{-1}=\left( 
\begin{array}{cc}
       0   & G(\x-\y)\\
-G(\x-\y) &  0
\end{array}
\right)    $$   $$    \eqno{(3.16)} $$
makes the brackets$\{\,,\,\}_3$ satisfy the
 usual bracket relations between $\pp,\pp^*,
p,p^*$.

$ A^1 $ and $ A^2 $ are now eliminated using strong equations
$$ \k(\vec \n \times \vec A)-\jj^0=0\eqno{(3.17)}$$
$$ \vec \n \cdot \vec A=0\eqno{(3.18)}$$
Their solution is
$$ A^1=-\p_2 B \quad,\quad A^2=\p_1 B \quad,\quad \n^2 B={\jj^0\oo\k}
\eqno{(3.19)}$$
therefore
$$A^1=-{1 \oo \k}\int d^2 y \p_{2x} G(\x-\y)\jj^0(y) \eqno{(3.20)}$$
$$A^2=+{1 \oo \k}\int d^2 y \p_{1x} G(\x-\y)\jj^0(y) \eqno{(3.21)}$$

If we integrate by parts assuming fast decrease of
$\jj_1,\jj_2$ at spatial infinity ,the expression for $ A^0\, $becomes
$$ A^0={-1\oo \k}\int d^2 y [\p_{1x}G (\x-\y) \jj_2-\p_{2x} G(\x-\y) \jj_1]
\eqno{(3.22)}$$
which makes it possible to write all three together ( as there  
is no time dependence
in $ G $);
$$ A^\mu={-1\oo \k}\int d^2 y \e^{\mu \nu \rho}[\p_\nu G(\x-\y)] \jj_\rho(y)
\eqno{(3.23)}$$
Thus all the degrees of freedom associated with $ A^\rho$ and their canonical
momenta $\pi_\rho$ are eliminated.What remains is a scalar field $\pp,\pp^*$
with $p,p^*$interacting with itself through a non-local interaction determined
by $\vec A$.

The reduced phase space Hamiltonian with Dirac bracket$\{\,,\,\}_3$ is
$$H=\int d^2 x[p^* p+(\vec D \pp)^*\cdot(\vec D \pp)+m^2 |\pp|^2]
\eqno{(3.24)}$$
with $\vec D= \vec \n + ie \vec A $.
The equations of motion in reduced phase space are determined using
formulas like
$$\{\pp(\x,t),\jj^0(\y,t)\}=-ie \pp(x) \dd(\x-\y)\eqno{(3.25)}$$
$$\{\pp(\x,t),\vec A(\y,t)\}={ie\oo\k} \pp(\x,t) \vec G(\x-\y)\eqno{(3.26)}$$
etc.
where we use $ \vec G=(G^1,G^2)=\e^{ij} \p_{jx} G(\x-\y) $
and $\vec \jj=(\jj_1,\jj_2) $.They are the same equations as (2.22-25)
except that now $A^\mu$ is expressed through (3.23).
\section{Dirac-Schwinger condition and  Poincare Generators}
The relativistic invariance of the original lagrangian is totally obscured
by guage fixed non-local Hamiltonian$(3.24)$
.For a local field system, the Hamiltonian
density $\Theta^{00}(\x,t)$ must satisfy the Dirac Schwinger consistency
condition with respect to the Dirac bracket ( our $\{\,,\,\}_3$
we omit the index $3$
now)
$$\{\Theta^{00}(\x,t),\Theta^{00}(\y,t)\}=-\p_i \dd(\x-\y)
\{\Theta^{0i}(\x,t)+\Theta^{0i}(\y,t)\}\eqno{(4.1)}$$

Our $\Theta^{00}$ satisfies this condition as can be verified through a
straightforward but tedious calculation.The result is
$$\Theta^{0i}(x)=p D^i \pp +p^*(D^i \pp)^*=
p\p^i \pp +p^* \p^i \pp^* +A^i \jj^0 \eqno{(4.2)}$$
The Poincare generators can now be constructed
$$P^0=\int d^2 x \Theta^{00} =H\eqno{(4.3)}$$
$$P^i=\int \Theta^{0i}d^2 x \eqno{(4.4)}$$
$$J^{\mu \nu}=\int (x^\mu\Theta^{0\nu}-x^\nu\Theta^{0\mu}) d^2 x\eqno{(4.5)}$$
the algebra of Poincare generators is gauranteed by the
Dirac-Schwinger conditions. 
Of special interest is the change in transformation properties of our
basic variables $\pp $ and $ A^\mu $.The space translation generators give
the expected normal transformation
$$\{\pp(x),P^\mu\}=\p^\mu \pp(x) \eqno{(4.6)}$$ with similar formulas
for $p,\pp^*,p^*$ and$ \jj_\mu$ follow  rather easily.
For $ A^\mu $ ,we see , for example 
\begin{eqnarray*}
\{A^1(x),P^2\}&=&{-1\oo \k}\int d^2 y \p_{2x} G(\x-\y)\{\jj^0(y),P^2\}  \\
              &=&{+1\oo \k}\int d^2 y \p_{2x} G(\x-\y)\p_{2y} \jj^0(y)\\
              &=&{-1\oo \k}\int d^2 y \p_{2y}\p_{2x} G(\x-\y)\jj^0(y)\\
             &=& +\p^2_x A^1(x)
             \end{eqnarray*}
             $$\eqno{( 4.7)}$$
As we are in the gauge fixed reduced phase
space modification of transformation properties is expected.The ``\,anomalous"
transformations are those under rotations and boosts.
The rotation generator is
\begin{eqnarray*}
 J^{12}&=&-\int d^2 y [y^1 (p \p_2 \pp+p^* \p_2 \pp^*)-y^2
(p \p_1 \pp+p^* \p_1 \pp^*)]\\
  & +  & \int d^2 y [y^1  A^2-y^2  A^1]  \jj_0
\end{eqnarray*} $$ \eqno{(4.8)}$$
The second integral in $ J^{12}  $
can be simplified to 
$$\int d^2 y [ y^1 A^2-y^2 A^1 ]\jj_0 ={1\oo{4\pi \k}}Q^2 \eqno{(4.9)}$$
where $Q=\int d^2 y \jj_0 $
using the identity
$$(\x-\y)\cdot\vec \n G(\x-\y)={1\oo 2\pi} \eqno{(4.10)}$$

Transformations under rotation for $\pp,p$ etc. contain extra infinitesimal
x-independent phase transformations
$$\e\{\pp(\x,t),M^{12}\}=\e(x^1
\p^2-x^2 \p^1 )\pp-{i e \e\oo 2\pi \k}Q \pp(\x,t)\eqno{(4.11)}$$
or for $\p x^\mu=(0,\e x^2,-\e x^1)$
$$\e\{\pp(x),J^{12}\}=\pp'(x)-\pp(x)=-\dd x^\mu\p_\mu \pp-i e \alpha \pp
\eqno{(4.12)}$$
gives $\pp'(x')=e^{-i e \alpha }\pp (x) $.
the extra term has been called in the the literature
the ``\,rotational anomaly".The transformations for $\jj_\mu$ are anomaly
free, as expected ,because they are gauge invariant.$$\{ \jj^0,J^{12}\}=0
\eqno{(4.13)}$$
$$\{ \jj_1,J^{12}\}=(x^1 \p^2-x^2 \p^1)\jj_1-\jj_2\eqno{(4.14)}$$
$$\{ \jj_2,J^{12}\}=(x^1 \p^2-x^2 \p^1)\jj_2-\jj_1\eqno{(4.15)}$$

These formulas determine the transformations for $A_\mu$
$$\{ A^0,J^{12}\}=(x^1 \p^2-x^2 \p^1)A^0\eqno{(4.16)}$$
$$\{A^1,J^{12}\}=(x^1 \p^2-x^2 \p^1)A^1-A^2\eqno{(4.17)}$$
$$\{A^2,J^{12}\}=(x^1 \p^2-x^2 \p^1)A^2+A^1\eqno{(4.18)}$$

The boost generators

$$J^{0i}=\int (y^0 \Theta^{0i}-y^i \Theta^{00}) d^2 y \eqno{(4.19)}$$
        lead to somewhat more involved transformation laws.
        The ``boost anomaly"
is no longer a constant phase transformation;
$$\e\{ \pp(x),J^{0i}\}=\e (x^0 \p^1-x^1 \p^0)\pp+i e \alpha^{(i)} (x) \pp(x)
\eqno{(4.20)}$$
$$\alpha^{(i)} (x)={ \e \oo \k}\int d^2 y (y^i-x^i)[(\p_1 G)(\x-\y) \jj_2-
(\p_2 G)(\x-\y) \jj_1]\eqno{(4.21)}$$
or $\pp'(x')=e^{i e \alpha^{(i)} (x)}\pp(x)  $
$$x'=x +\dd x\qquad,\qquad \dd x^\mu=(\e x^0,\e x^1,0)$$
with similar formulas for $p$ etc.
The charge density $\jj^0 $ is anomaly free
$$ \e \{\jj_0,J^{01}\}=-\dd x^\mu \p_\mu \jj_0-\e \jj_1 \eqno{(4.22)}$$
and $A^1$ for example gives after frequent integration by parts to juggle
derivatives and using $\p_0 \jj_0 =\p_i \jj_i $
$$\e \{ A^1, J^{01}\}=-\dd x^\mu \p_\mu A^1+\e A^0+\p^1 \alpha^{(i)} (x)
\eqno{(4.23)}$$
   which shows that $A^\mu $transforms as a ($2+1$) vector but upto a gauge
transformation with the compensating $\alpha^{(i)} (x) $ determined by
 the boost anomaly in $\pp(x)$.This proves that all gauge invariant quantities
 will have normal transformation because of correct transformation properties
 of $\pp $ and $ A^\mu $.  The verification of Poincare algebra
 is straightforward , if a little tedious , but has been done.

 \section{Discussion}       
As far as we are aware, the gauge conditions for this system have not
been properly chosen earlier notwithstanding the
 fact the reduced phase space
Hamiltonian and Canonical brackets are the same as obtained earlier
by other authors.The question that naturally arises is that in what way
has the proper choice of gauge conditions altered the dynamics of the system?

The answer lies in the Lorentz boost transformation properties of
physical quantities.For example,the boost transformation shifts the time
component like in the equation for $ A^1 $ above.
With wrong gauge conditions (for example $A^0=0$) such a transformation
 law cannot be obtained.\\Therefore , our system is truely frame independent
 and any Lorentz transformation transforms the basic fields $\pp$ with
 a phase factor $\alpha $ which is exactly compensated by the corresponding
 $\p_\mu \alpha $ in $A^\mu $.
 therefore all physically relevant quantities such as $\jj_\mu$ which are
 gauge invariant have proper transformation laws without anomalies.

 The intimate connection between consistency
  of gauge conditions with equations of
 motion and Lorentz transformation laws has probably been missed before
 because previous authors have analysed transformations under rotations
 but not,to our knowledge,under boosts.$^7$
 It is only the boosts which involve time development ,and therefore
 implicitly use equations of motion.
We hope that with this analysis we have not only suggested
the correct gauge conditions ,but also pointed out the relevance
 of transformations under boosts for constrained systems.

\section{Acknowledgement}
Pankaj Sharan is grateful to A.P.Balachandran,R.Rajaraman,
A.M. Srivastava and P.K.Panigrahi for fruitful discussions.\\
K.Dasgupta acknowledges the financial support by C.S.I.R.\\(Award No.
9/466(30)/96-EMR-I).\\
We are all grateful to  P.P.Singh for technical help in preparing this manuscript.
\section{References}
 1. C.Hagen,Ann.Phys.(N.Y.){\bf 157},342(1984)\\
 2.  G.Semenoff ,Phy.Rev.Lett. {\bf 61},517(1988);\\
\ \  G.Semenoff ,Phy.Rev.Lett. {\bf  63},1026(1989)   ;\\
\ \  G.Semenoff and P.Sodano,Nucl.Phys.{\bf B328},753(1989)\\
 3.  R.Banerjee Phy.Rev. D {\bf 48},2905(1993)\\
 4. A.Forester and H.O. Girotti Phys. lett.  {\bf B 230},83(1989)\\
 5. S.Forte , Rev.Mod.Phys.{\bf 64},193(1993)\\
6. For a review see F.Wilczek, ed., {\em Fractional Statistics and
Anyon Superconductivity} ( World Scietific , Singapore 1990 ).\\
7. Boosts are considered by C.Hagen in ref.1 , but not in the reduced
phase space with Dirac brackets.
\end{document}